# Optimization of allotropic and hardness of aluminum oxide coating by PEO

Babak Ghorbanian[a]*, Mohammad Tajally[a], Seyed Mohammad Mousavi Khoie[b], Hossien Tavakoli[a]

[a] Department of Metallurgical Engineering, Semnan University, Semnan, Iran.

[b] Department of Mining and Metallurgical Engineering, Amirkabir University of technology, Tehran Iran

*m_tajally@ semnan.ac.ir (M. Tajally)

**Abstract**

One of the most important methods of producing materials and oxide coating is plasma electrolytic oxidation (PEO). Coatings made on aluminum in PEO method have two allotropes of $\alpha$-$Al_2O_3$ and $\gamma$-$Al_2O_3$. Coatings containing $\alpha$-$Al_2O_3$ have more hardness and wear resistance; therefore, the main goal of the present study is to optimize aluminum oxide allotrope for increasing the ratio of $\alpha$-$Al_2O_3$ in coating made with PEO method. Results of the present study shows that the optimum electrolyte compound for making highest phase of $\alpha$-$Al_2O_3$ has 2.9 g/lit of KOH, 1.15 g/lit of $Na_4P_2O_7$, and 0.34 g/lit of $NaAlO_2$ and ratio of highest allotropic pick of $\alpha$-$Al_2O_3$ to highest allotropic pick of $\gamma$-$Al_2O_3$ (in XRD test) in the optimum condition is 0.622 and hardness is 1648 Vickers.

**Introduction**

According to features like high wear resistance and large weights, aluminum has wide applications in industries such as aerospace and automobile manufacturing. However, despite mentioned advantages, defects such as low hardness, low wear resistance and high friction coefficient has resulted in limitations for using aluminum alloys [1,2].

Plasma electrolytic oxidation (PEO) process is a novel method employed in the surface modification of aluminum samples. In this process, an oxide layer is prepared as the protector over the metal due to oxidation in the electrolyte, the characteristics of which are affected by the electrolyte composition [3, 4]. The results of various studies have demonstrated that during the process of electrolytic plasma, two allotropes of $\alpha$-$Al_2O_3$ and $\gamma$-$Al_2O_3$ are made that amount of these allotropes depend on parameters such as electrolyte feature, current, voltage and etc. Amount of $\gamma$-$Al_2O_3$ allotrope is always more than the other one [5]. The studies show that $\alpha$-$Al_2O_3$ has good mechanical properties (like hardness and wear resistance) and $\gamma$-$Al_2O_3$ has good corrosion properties. Therefore, the best coverage should include an optimal amount of both of the two allotropes [6].

This research is aimed at finding a proper candidate to enhance PEO treated aluminum's hardness by optimizing two allotropes at the same time. To do so the effect of changing amount of electrolyte components on PEO treated aluminum's characteristics is examined. In this study was used Response surface methodology (RSM) for optimizing hardness and allotropes.

**Material and Methods**

**PEO treatments**

The PEO process was carried out using a DC power source and it was performed at constant Voltage difference ($\Delta V = V_{final} - V_{Starting\ the\ plasma}$) 50 v, process time 10 min, A polymer reservoir as electrolyte keeper, stirring and cooling system [7-8]. The polished 1050 aluminum sheets were used as the working electrode and stainless steel as the counter electrode. Aqueous electrolyte was prepared using solution of 0-6 g/l $NaAlO_2$ (Merck), 0-3 g/l KOH (Aldrich) and 0-3 g/l $Na_4P_2O_7$ (Merck) in 2 liters distilled, pH=11.84.

**Characterization and Analysis**

Phase composition of coatings and bare alloy were studied by X-ray diffraction (XRD, Digaku D/max−2500) using Cu Kα radiation at 40 kV and 100 mA. In addition, the hardness of samples was checked using hardness tester machine (Koopa UV1) with a force of 5 grams.

**Response surface methodology**

In RSM, in order to determine code, Box-Behnken design (BBD) was used. In Box-Behnken statistical design using Design-Expert software [7], the range of changes for $NaAlO_2$, KOH, and $Na_4P_2O_7$ was 0-6 g/lit, 0-3 g/lit, and 0-3 g/lit, respectively. Dependent variables were the ratio of the highest allotropic pick of α-$Al_2O_3$ to highest allotropic pick of γ-$Al_2O_3$ ($\frac{I_\alpha}{I_\gamma}$) and hardness (Hv).

According to BBD, having 3 factors, we had to carry out 17 tests in order to study the effects of the 3 independent factors on two responses.

**Results and discussion**

The condition of the 17 tests along with results of dependent variables ($\frac{I_\alpha}{I_\gamma}$) and Hv of the samples have been shown in Table 1.

**Table 1.** test design using RMS along with results of XRD and hardness.

|   | X1(g/l) | X2(g/l) | X3(g/l) | $\frac{I_\alpha}{I_\gamma}$ | Hardness (Vickers) |
|---|---|---|---|---|---|
| 1 | 0 | 1.5 | 3 | 0.0301 | 801 |
| 2 | 1.5 | 1.5 | 3 | 0.1833 | 1008 |
| 3 | 0.608095 | 2.391905 | 1.216189 | 0.0283 | 799 |
| 4 | 0.608095 | 0.608095 | 1.216189 | 0.0262 | 798 |
| 5 | 3 | 1.5 | 3 | 0.5781 | 1620 |
| 6 | 0.608095 | 2.391905 | 4.783811 | 0.0358 | 809 |
| 7 | 1.5 | 1.5 | 3 | 0.1325 | 925 |
| 8 | 2.391905 | 0.608095 | 1.216189 | 0.4912 | 1550 |
| 9 | 1.5 | 3 | 3 | 0.024 | 782 |
| 10 | 0.608095 | 0.608095 | 4.783811 | 0.015 | 738 |
| 11 | 2.391905 | 2.391905 | 1.216189 | 0.0779 | 802 |
| 12 | 1.5 | 1.5 | 3 | 0.1084 | 841 |
| 13 | 2.391905 | 2.391905 | 4.783811 | 0.0106 | 725 |
| 14 | 2.391905 | 0.608095 | 4.783811 | 0.3821 | 1410 |
| 15 | 1.5 | 0 | 3 | 0.0967 | 825 |
| 16 | 1.5 | 1.5 | 3 | 0.1431 | 965 |

| 17 | 1.5 | 1.5 | 0 | 0.1422 | 952 |
| 18 | 1.5 | 1.5 | 6 | 0.1311 | 912 |

**Model of ratio of different allotropic picks ($\frac{I_\alpha}{I_\gamma}$)**

Obtaining R2, R2-Adjusted (0.9598-0.8864) and P-value (0.0013<0.05) in fitting models shows that models of RSM are ideal. F-value obtained from response coefficients (dependent variable $\frac{I_\alpha}{I_\gamma}$) showed that linear coefficients X1 (F-value=51.71 and p-value=0.0001) and X2 (F-value=12.53 and p-value=0.0076), quadratic coefficient $X1^2$ (F-value=6.95 and p-value=0.0299), and coefficient of reciprocal effects of X1X2 (F-value=18.15 and p-value=0.0028) were significant 5% in case of probability. In fact, results of analyzing variance with the expected model for dependent variable $\frac{I_\alpha}{I_\gamma}$ was significant. Accordingly, fitted models for the obtained response of $\frac{I_\alpha}{I_\gamma}$ are as follows:

$$\frac{I_\alpha}{I_\gamma} = -0.27426 + 0.20329X_1 + 0.24804X_2 + 0.022069X_3 - 0.13472X_1X_2 - 0.014342X_1X_3$$
$$+ 0.00499167X_2X_3 + 0.063596X_1^2 - 0.044737X_2^2 - 0.00270648X_3^2$$
$$(1)$$

That $\frac{I_\alpha}{I_\gamma}$ is the *expected response* for the ratio of the highest allotropic pick of $\alpha$-Al$_2$O$_3$ to highest allotropic pick of $\gamma$-Al$_2$O$_3$ and $X_1$, $X_2$, and $X_3$ are coded amounts for KOH, Na$_4$P$_2$O$_7$, and NaAlO$_2$, respectively. Regression analysis of data showed that in case of presence of all three factors, amount of Na$_4$P$_2$O$_7$ has the most effect on the ratio of the highest allotropic pick of $\alpha$-Al$_2$O$_3$ to highest allotropic pick of $\gamma$-Al$_2$O$_3$; i.e. the amount of Na$_4$P$_2$O$_7$ controls the amount of $\alpha$-Al$_2$O$_3$. on the other hand, the smallest effect was for NaAlO$_2$.

3D diagram of $\frac{I_\alpha}{I_\gamma}$ according to X1-X3 variables has been shown in Fig.1 The highest amount of each diagram is about optimum effects of the two independent factors on the respective parameter. Accordingly, in Fig.1 (A) that is about the reciprocal effect of X1 and X2, in the constant amount of aluminate, the ratio of $\frac{I_\alpha}{I_\gamma}$ is increased by increasing amount of KOH and the highest amount of $\frac{I_\alpha}{I_\gamma}$ is obtained in 2.37 g/lit of KOH. Fig.1 (B) shows reciprocal effect of X1 and X3. According to this figure, increasing amount of KOH controls the ratio of $\frac{I_\alpha}{I_\gamma}$ and amount of aluminate has no effect on the allotropes of the produced alumina. Fig.1 (C) shows the reciprocal effect of X2 and X3. Data obtained from this part shows that reciprocal effect of NaAlO$_2$ and Na$_4$P$_2$O$_7$ does not control the ratio of $\frac{I_\alpha}{I_\gamma}$, because the obtained amounts in the constant amount of KOH are very close each other. Therefore, results show that in case of reciprocal effects, KOH is a controller for the ratio of $\frac{I_\alpha}{I_\gamma}$.

Design-Expert® Software

R1
0.5781
0.0106

X1 = A: KOH
X2 = B: phosphat

Actual Factor
C: aluminat = 3.00

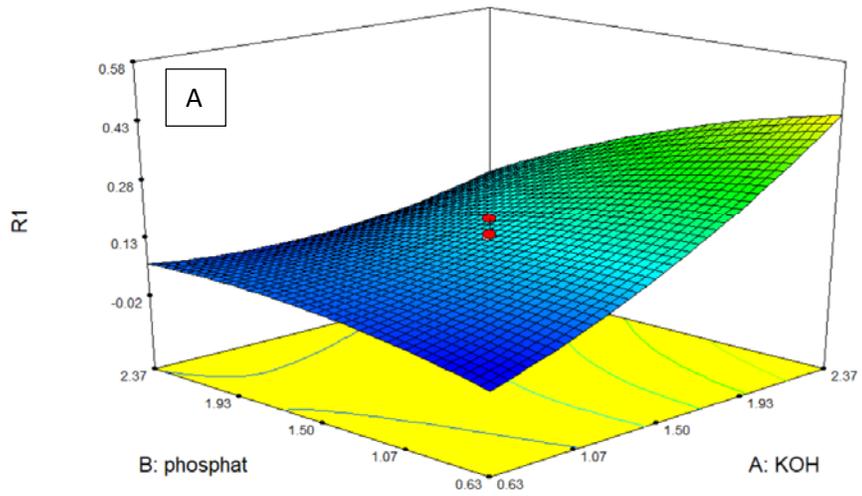

R1
0.5781
0.0106

X1 = A: KOH
X2 = C: aluminat

Actual Factor
B: phosphat = 1.50

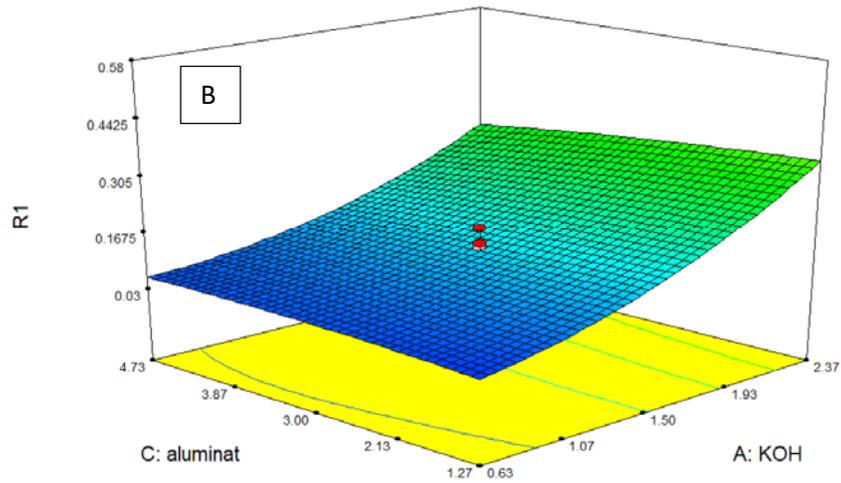

Design-Expert® Software

R1
0.5781
0.0106

X1 = B: phosphat
X2 = C: aluminat

Actual Factor
A: KOH = 1.50

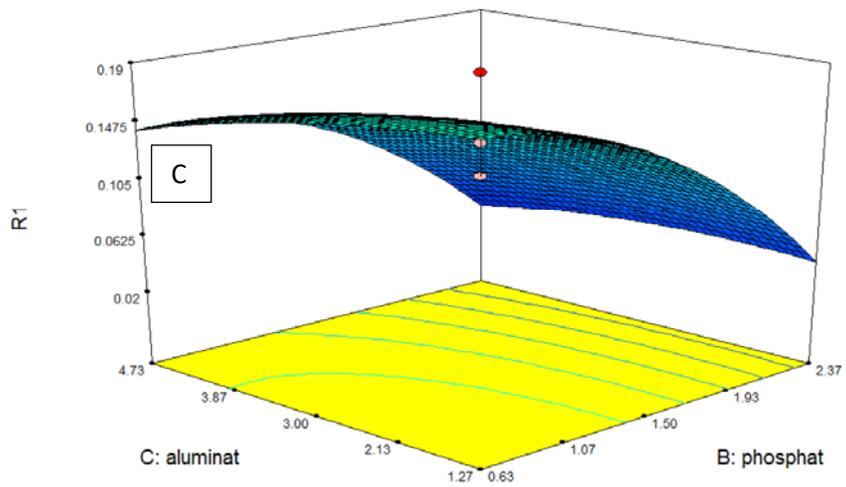

**Fig. 1.** reciprocal effect of independent variables on dependent variable of $\frac{I_\alpha}{I_\gamma}$. (A) X1 and X2, (B) X1 and X3, and (C) X2 and X3

**Hardness model**

Obtaining R2, R2-Adjusted (0.9621-0.9121) and P-value (0.003<0.05) in fitting models shows that models of RSM are ideal. F-value obtained from response coefficient (dependent variable of hardness) showed that linear coefficients X1 (F-value=36.04 and p-value=0.0006) and X2 (F-value=9.74 and p-value=0.0124), quadratic coefficient $X1^2$ (F-value=18.73 and p-value=0.0025), and coefficient of reciprocal effects of X1X2 (F-value=6.81 and p-value=0.0311) were significant 5% in case of probability. In fact, results of analyzing variance with expected models from response surface methodology showed that the hardness model was significant for every response (F-value=8.52 and p-value=0.003). Accordingly, fitted models of hardness are as follows:

$$Hv = 328.70603 + 299.53068X_1 + 421.03233X_2 + 2.92958X_3 - 250.83333X_1X_2 \\ - 13.91667X_1X_3 + 11.08333X_2X_3 + 115.41270X_1^2 - 65.47619X_2^2 \\ - 2.09127X_3^2$$

(2)

Regression analysis of data showed that in case of presence of all three factors, the amount of $Na_4P_2O_7$ has the most effect on hardness; i.e. the amount of $Na_4P_2O_7$ controls the amount of hardness. On the other hand, the smallest effect was for $NaAlO_2$. These results have the correlation with obtained results for $\frac{I_\alpha}{I_\gamma}$. Therefore, according to obtained results from hardness model and $\frac{I_\alpha}{I_\gamma}$ model, it can be concluded that increasing amount of allotropic pick of $\alpha$-$Al_2O_3$, directly affects hardness and wear. Therefore, one of the most important achievements of the present study is that by optimizing allotropic coating by increasing $\alpha$-$Al_2O_3$, we can increase hardness.

Fig. 2 shows the 3D diagram of reciprocal effects of factors X1 to X3 on hardness. As it can be seen in Fig. 2 (A), in case of the reciprocal effect of KOH and $Na_4P_2O_7$ on hardness, hardness increases by increasing amount of KOH, which is because of the effect of KOH on increasing $\alpha$-$Al_2O_3$. In addition, in case of the reciprocal effect of $NaAlO_2$ and KOH on hardness, again the effect of KOH is more, which is because of increasing $\alpha$-$Al_2O_3$ in result of increasing KOH (Fig. 2(B)). Fig. 2 (C) shows the reciprocal effect of X2 and X3. Data obtained from this part shows that the reciprocal effect of $NaAlO_2$ and $Na_4P_2O_7$ does not control the hardness, because the obtained amounts in the constant amount of KOH are very close to each other. Therefore, results show that in case of reciprocal effects, KOH is a controller for the ratio of hardness.

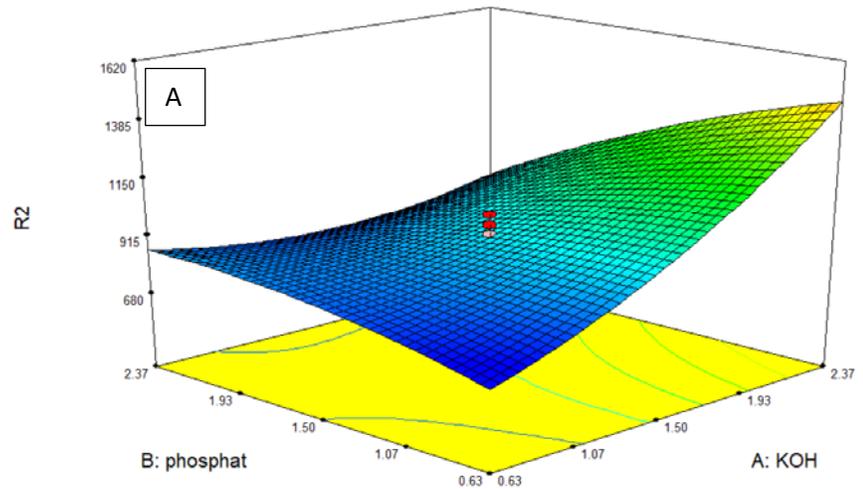

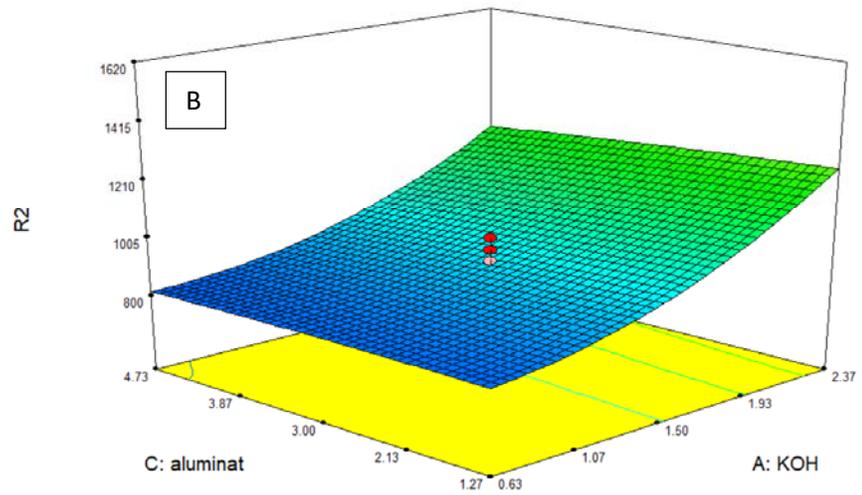

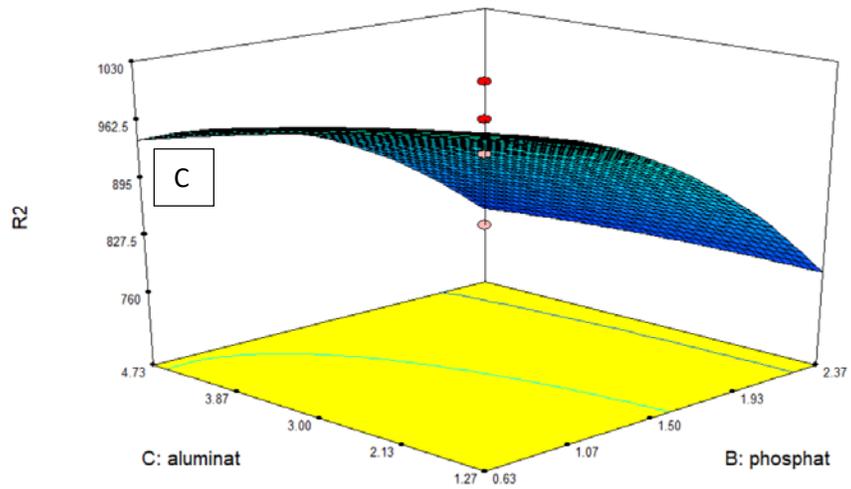

**Fig. 2.** reciprocal effect of independent variables on dependent variable of hardness. (A) X1 and X2, (B) X1 and X3, and (C) X2 and X3

**Simultaneous optimization of variables**

Fig. 3 (A) is a 3D diagram for changing amounts of KOH and sodium phosphate according to the percentage of obtaining the optimum sample. According to this diagram, the most probable optimum condition is with 2.9 g/lit of KOH, 1.15 g/lit of $Na_4P_2O_7$, and 1.15 g/lit of $NaAlO_2$. In this condition, optimum output results show that the ratio of $\frac{I_\alpha}{I_\gamma}$ is 0.622 and amount of hardness is 1648 Vickers.

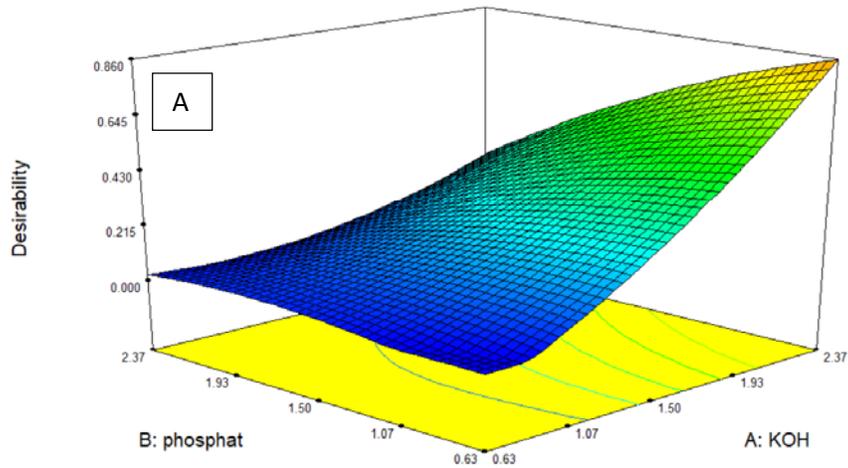

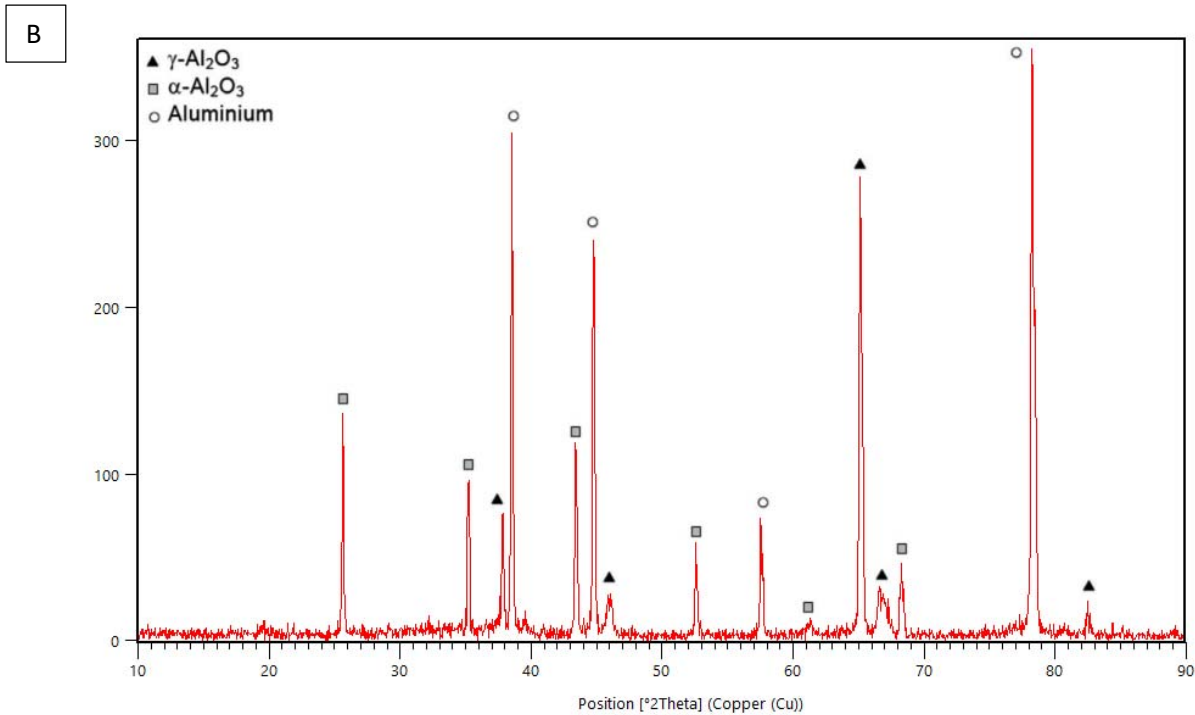

**Fig.3.** (A) 3D diagram of changing of KOH and $Na_4P_2O_7$ according to probability of achieving. (B) XRD for optimal sample

Fig. 3(B) is a sample of XRD patterns obtained from simple XRD for coatings made with optimal electrolytes. Results obtained from XRD showed that in optimal condition, $\frac{I_\alpha}{I_\gamma}$ is 0.62, which has lower than 5% of difference with the amount obtained in statistical simulation (0.622). In addition, the amount of obtained hardness is 1651 Vickers, which has lower than 5% of difference with output results of the software.

**Conclusion**

In the optimal condition, $\frac{I_\alpha}{I_\gamma}$ equals to 0.62, which has lower than 5% of the difference from amount obtained in the statistical simulation (0.622); therefore, it can be said that results of the designation have a correlation with results of the fact-checking. The obtained hardness amount is 1651, which this amount has lower than 5% of the difference from output results of the software. Therefore, this approves fact-checking of results of this study.